\documentclass[11pt]{article}
\usepackage{amssymb,amsmath,amsfonts}
\usepackage{graphicx}
\usepackage{graphics}
\usepackage{eepic,epsfig}

\begin{document}

\title{Induced fermionic current in toroidally compactified spacetimes with
applications to cylindrical and toroidal nanotubes}
\author{ S. Bellucci$^{1}$\thanks{%
E-mail: bellucci@lnf.infn.it }, A. A. Saharian$^{2}$\thanks{%
E-mail: saharian@ysu.am }, V. M. Bardeghyan$^{2}$ \vspace{0.3cm} \\
\textit{$^1$ INFN, Laboratori Nazionali di Frascati,}\\
\textit{Via Enrico Fermi 40,00044 Frascati, Italy} \vspace{0.3cm}\\
\textit{$^2$ Department of Physics, Yerevan State University,}\\
\textit{1 Alex Manoogian Street, 0025 Yerevan, Armenia }}
\maketitle

\begin{abstract}
The vacuum expectation value of fermionic current is evaluated for a massive
spinor field in spacetimes with arbitrary number of toroidally compactified
spatial dimensions in the presence of a constant gauge field. By using the
Abel-Plana type summation formula and the zeta function technique we present
the fermionic current in two different forms. Non-trivial topology of the
background spacetime leads to the Aharonov-Bohm effect for the fermionic
current induced by the gauge field. The current is a periodic function of
the magnetic flux with the period equal to the flux quantum. In the absence
of gauge field it vanishes for special cases of untwisted and twisted
fields. Applications of general formulae to Kaluza-Klein type models and to
cylindrical and toroidal carbon nanotubes are given. In the absence of
magnetic flux the total fermionic current in carbon nanotubes vanishes, due
to the cancellation of contributions from two different sublattices of the
hexagonal lattice of graphene.
\end{abstract}

\bigskip


\bigskip

\section{Introduction}

In many physical problems we need to consider some model on a background of
manifold having compact spatial dimensions along which dynamical variables
satisfy some prescribed periodicity conditions. Incomplete list of
applications where the topological effects play an important role includes
Kaluza-Klein type models, supergravity and superstring theories. From an
inflationary point of view, universes with compact dimensions, under certain
conditions, should be considered as a general rule rather than an exception
\cite{Lind04}. Models of a compact universe with non-trivial topology may
play an important role by providing proper initial conditions for inflation.
An interesting application of the field theoretical models with non-trivial
topology of spatial dimensions appeared in nanophysics recently \cite{Sait98}%
. The long-wavelength description of the electronic states in graphene can
be formulated in terms of Dirac-like theory in 3-dimensional spacetime with
the Fermi velocity playing the role of a speed of light (see, e.g., Refs.
\cite{Seme84,Vinc84}). Single-walled carbon nanotubes are generated by
rolling up a graphene sheet to form a cylinder and the background spacetime
for the corresponding Dirac-like theory has a topology $R^{2}\times S^{1}$.
The compactification in the direction along the cylinder axis gives another
class of graphene structures called toroidal carbon nanotubes with the
background topology $R^{1}\times (S^{1})^{2}$ \cite{Liu97}.

The compactification of spatial dimensions leads to a number of interesting
field theoretical effects which include instabilities in interacting field
theories, topological mass generation and symmetry breaking. In quantum
field theory the boundary conditions imposed on fields along compact
dimensions change the spectrum of vacuum fluctuations. The resulting
energies and stresses are known as the topological Casimir effect. (For the
topological Casimir effect and its role in cosmology see \cite{Most97}-\cite%
{Duff86} and references therein.) Note that Casimir forces between material
boundaries are presently attracting much experimental attention \cite{Klim09}%
. In the Kaluza-Klein type models this effect has been used as a
stabilization mechanism for moduli fields which parametrize the size and the
shape of the extra dimensions. The Casimir energy can also serve as a model
of dark energy needed for the explanation of the present accelerated
expansion of the universe (see \cite{Milt03} and references therein).

The effects of the toroidal compactification of spatial dimensions on the
properties of quantum vacuum for various spin fields have been discussed by
several authors (see, for instance, \cite{Most97}-\cite{CasTor} and
references therein). One-loop quantum effects for the scalar and fermionic
fields in de Sitter spacetime with toroidally compactified dimensions are
studied in Refs. \cite{Saha08,Saha08f}. In previous papers \cite%
{Bell09a,Bell09b} we have investigated the fermionic condensate
and the vacuum expectation value of the energy-momentum tensor for
a massive spinor field in higher-dimensional spacetimes with
toroidally compactified spatial dimensions. These expectation
values are among the most important quantities that characterize
the properties of the quantum vacuum. Another important
characteristic is the vacuum expectation value of fermionic
current. Although the corresponding operator is local, due to the
global nature of the vacuum, this quantity carries an important
information about the global properties of the background
spacetime. In addition to describing physical structure of the
quantum field at a given point, the current acts as the source in
the Maxwell equations. It therefore plays an important role in
modelling a self-consistent dynamics involving the electromagnetic
field.

In the present paper, we investigate one-loop quantum effects on
the fermionic current arising from vacuum fluctuations of a
massive fermionic field on the background of spacetimes with an
arbitrary number of toroidally compactified spatial dimensions. We
will assume generalized periodicity conditions along the
compactified dimensions with arbitrary phases and the presence of
a constant gauge field. A non-zero gauge field defined on
topologically non-trivial background leads to the Aharonov-Bohm
effect for the vacuum expectation value of fermionic current. Note
that fermionic current in spacetime with non-trivial topology
induced by a cosmic string has been investigated in Refs.
\cite{Beze94CosStr}.

This paper is organized as follows. In the next section, we consider vacuum
expectation value of the fermionic current in the background spacetime with
spatial topology $R^{p}\times (S^{1})^{q}$ in the presence of a constant
gauge field. The corresponding expression is derived by using the Abel-Plana
type summation formula. An equivalent representation is obtained in Section %
\ref{sec:FCZeta} within the framework of the generalized zeta function
approach. In Section \ref{sec:FCNano} we apply the general formula for the
evaluation of the fermionic current in cylindrical and toroidal nanotubes
within the framework of Dirac-like model for electrons in graphene. Main
results are summarized in Section \ref{sec:Conc}.

\section{Vacuum expectation value of the fermionic current}

\label{sec:FC}

We consider the quantum fermionic field $\psi $ on a background of $(D+1)$%
-dimensional flat spacetime with spatial topology $R^{p}\times (S^{1})^{q}$,
$p+q=D$. The Cartesian coordinates along uncompactified and compactified
dimensions are denoted as $\mathbf{z}_{p}=(z^{1},\ldots ,z^{p})$ and $%
\mathbf{z}_{q}=(z^{p+1},\ldots ,z^{D})$, respectively. The length of the $l$%
-th compact dimension we denote as $L_{l}$. Hence, for coordinates one has $%
-\infty <z^{l}<\infty $ for $l=1,\ldots ,p$, and $0\leqslant z^{l}\leqslant
L_{l}$ for $l=p+1,\ldots ,D$. We assume that along the compact dimensions
the field obeys the generic quasiperiodic boundary conditions,%
\begin{equation}
\psi (t,\mathbf{z}_{p},\mathbf{z}_{q}+L_{l}\mathbf{e}_{l})=e^{2\pi i\alpha
_{l}}\psi (t,\mathbf{z}_{p},\mathbf{z}_{q}),  \label{BC1}
\end{equation}%
with constant phases $\alpha _{l}$ and with $\mathbf{e}_{l}$\ being the unit
vector along the direction of the coordinate $z^{l}$, $l=p+1,\ldots ,D$.
Condition (\ref{BC1}) includes the periodicity conditions for both untwisted
and twisted fermionic fields as special cases with $\alpha _{l}=0$ and $%
\alpha _{l}=1/2$, respectively. As is discussed below, the special cases $%
\alpha _{l}=0,\pm 1/3$ are realized in nanotubes.

Dynamics of the massive spinor field is governed by the Dirac equation
\begin{equation}
i\gamma ^{\mu }D_{\mu }\psi -m\psi =0\ ,\;D_{\mu }=\partial _{\mu }+ieA_{\mu
},  \label{Direq}
\end{equation}%
where $A_{\mu }$ is the vector potential for the external electromagnetic
field. In the discussion below we assume that $A_{\mu }=\mathrm{const}$.
Though the corresponding magnetic field strength vanishes, the non-trivial
topology of the background spacetime leads to Aharonov-Bohm-like effects for
physical observables. In particular, as it is shown below, the expectation
value of fermionic current depends on $A_{\mu }$. In the $(D+1)$%
-dimensional spacetime, the Dirac matrices are $N\times N$ matrices with $%
N=2^{[(D+1)/2]}$, where the square brackets mean the integer part
of the enclosed expression. We assume that these matrices are
given in the Dirac representation:
\begin{equation}
\gamma ^{0}=\left(
\begin{array}{cc}
1 & 0 \\
0 & -1%
\end{array}%
\right) ,\;\gamma ^{\mu }=\left(
\begin{array}{cc}
0 & \sigma _{\mu } \\
-\sigma _{\mu }^{+} & 0%
\end{array}%
\right) ,\;\mu =1,2,\ldots ,D.  \label{DiracMat}
\end{equation}%
From the anticommutation relations for the Dirac matrices one has $\sigma
_{\mu }\sigma _{\nu }^{+}+\sigma _{\nu }\sigma _{\mu }^{+}=2\delta _{\mu \nu
}$. In the case $D=2$ we have $N=2$ and the Dirac matrices are taken in the
form $\gamma ^{\mu }=(\sigma _{\text{P}3},i\sigma _{\text{P}1},i\sigma _{%
\text{P}2})$, with $\sigma _{\text{P}\mu }$ being the $2\times 2$ Pauli
matrices. We are interested in the effects of non-trivial topology on the
vacuum expectation value (VEV) of fermionic current $j^{\mu }=\bar{\psi}%
\gamma ^{\mu }\psi $, where $\bar{\psi}$ is the Dirac conjugated spinor.
Note that the fermionic condensate and the VEV of the energy-momentum tensor
in the model under consideration were evaluated in Ref. \cite{Bell09a} in
the absence of gauge field.

By expanding the field operator in terms of annihilation and
creation operators, the VEV of fermionic current is presented as
the sum over
all modes%
\begin{equation}
\langle 0|j^{\mu }|0\rangle =\sum_{\beta }\bar{\psi}_{\beta }^{(-)}(x)\gamma
^{\mu }\psi _{\beta }^{(-)}(x),  \label{fcvev}
\end{equation}%
where $\{\psi _{\beta }^{(+)},\psi _{\beta }^{(-)}\}$ is the
complete set of positive- and negative-frequency eigenfunctions
satisfying the periodicity conditions (\ref{BC1}) along compact
dimensions. Here, $\beta $ is a set of quantum numbers specifying
the solutions. The dependence of the
eigenfunctions on the spacetime coordinates can be taken in the form $e^{i%
\mathbf{k}\cdot \mathbf{r}-i\omega t}$, with the wave vector $\mathbf{k}$.
From the Dirac equation for the positive- and negative-frequency solutions
we find%
\begin{eqnarray}
\psi _{\beta }^{(+)} &=&A_{\beta }^{(+)}e^{i\mathbf{k}\cdot \mathbf{r}%
-i\omega t}\left(
\begin{array}{c}
w_{\sigma }^{(+)} \\
\frac{\omega -eA_{0}-m}{(\mathbf{k}-e\mathbf{A})^{2}}(\mathbf{k}-e\mathbf{A}%
)\cdot \boldsymbol{\sigma}^{+}w_{\sigma }^{(+)}%
\end{array}%
\right) ,  \label{psibet+} \\
\psi _{\beta }^{(-)} &=&A_{\beta }^{(-)}e^{-i\mathbf{k}\cdot \mathbf{r}%
+i\omega t}\left(
\begin{array}{c}
\frac{\omega +eA_{0}-m}{(\mathbf{k}+e\mathbf{A})^{2}}(\mathbf{k}+e\mathbf{A}%
)\cdot \boldsymbol{\sigma}w_{\sigma }^{(-)} \\
w_{\sigma }^{(-)}%
\end{array}%
\right) ,  \label{psibet-}
\end{eqnarray}%
where $\beta =(\mathbf{k},\sigma )$, $A_{\mu }=(A_{0},-\mathbf{A})$, and $%
\boldsymbol{\sigma}=(\sigma _{1},\sigma _{2},\ldots ,\sigma _{D})$. In these
expressions $w_{\sigma }^{(+)}$, $\sigma =1,\ldots ,N/2$, are one-column
matrices having $N/2$ rows with the elements $w_{\sigma l}^{(+)}=\delta
_{\sigma l}$, and $w_{\sigma }^{(-)}=iw_{\sigma }^{(+)}$. The frequency and
the wave vector are connected by the relation $\left( \omega \mp
eA_{0}\right) ^{2}=(\mathbf{k}\mp e\mathbf{A})^{2}+m^{2}$ for the function $%
\psi _{\beta }^{(\pm )}$.

The coefficients $A_{\beta }^{(\pm )}$ are found from the orthonormalization
condition $\int d^{3}x\,\psi _{\beta }^{(\pm )+}\psi _{\beta ^{\prime
}}^{(\pm )}=\delta _{\beta \beta ^{\prime }}$, where $\delta _{\beta \beta
^{\prime }}$ is understood as the Dirac delta function for continuous
indices and the Kronecker delta for discrete ones. From this condition one
finds%
\begin{equation}
A_{\beta }^{(\pm )2}=\frac{1}{(2\pi )^{p}V_{q}}\left[ 1+\frac{\left( \omega
\mp eA_{0}-m\right) ^{2}}{(\mathbf{k}\mp e\mathbf{A})^{2}}\right] ^{-1},
\label{Abetpm}
\end{equation}%
where $V_{q}=L_{p+1}\cdots L_{D}$ is the volume of the compact subspace.

We decompose the wave vector into components along the uncompactified and
compactified dimensions: $\mathbf{k}=(\mathbf{k}_{p},\mathbf{k}_{q})$, $k=%
\sqrt{\mathbf{k}_{p}^{2}+\mathbf{k}_{q}^{2}}$. The eigenvalues for the
components along the compact dimensions are determined from boundary
conditions (\ref{BC1}):%
\begin{equation}
k_{l}=2\pi (n_{l}+\alpha _{l})/L_{l},\;n_{l}=0,\pm 1,\pm 2,\ldots
,\;l=p+1,\ldots ,D.  \label{kDn}
\end{equation}%
For the components along the uncompactified dimensions one has $-\infty
<k_{l}<\infty $, $l=1,\ldots ,p$.

Substituting the eigenfunctions (\ref{psibet-}) into the mode sum formula (%
\ref{fcvev}) and by using the properties of Dirac matrices one finds%
\begin{eqnarray}
\langle 0|j^{0}|0\rangle  &=&\frac{N}{2V_{q}}\int \frac{d\mathbf{k}_{p}}{%
(2\pi )^{p}}\sum_{\mathbf{n}_{q}\in \mathbf{Z}^{q}}1,  \label{j0} \\
\langle 0|j^{l}|0\rangle  &=&\frac{N}{2V_{q}}\int \frac{d\mathbf{k}_{p}}{%
(2\pi )^{p}}\sum_{\mathbf{n}_{q}\in \mathbf{Z}^{q}}\frac{(\mathbf{k}+e%
\mathbf{A})_{l}}{\sqrt{(\mathbf{k}+e\mathbf{A})^{2}+m^{2}}},  \label{jl}
\end{eqnarray}%
with $l=1,2,\ldots ,D$ and $\mathbf{n}_{q}=(n_{p+1},\ldots
,n_{D})$. In order to give a meaning to divergent expressions, it
is necessary to regularize them. Here we use a Pauli-Villars
gauge-invariant regularization. An alternative way is to introduce
a cutoff function. Introducing regulator fields with large masses
$M_{s}$, $s=1,2,\ldots ,S$,
for the regularized expressions one finds%
\begin{eqnarray}
\langle 0|j^{0}|0\rangle _{\text{reg}} &=&\frac{N}{2V_{q}}\int \frac{d%
\mathbf{k}_{p}}{(2\pi )^{p}}\sum_{\mathbf{n}_{q}\in \mathbf{Z}%
^{q}}\sum_{s=0}^{S}C_{s},  \label{j0reg} \\
\langle 0|j^{l}|0\rangle _{\text{reg}} &=&\frac{N}{2V_{q}}\int \frac{d%
\mathbf{k}_{p}}{(2\pi )^{p}}\sum_{\mathbf{n}_{q}\in \mathbf{Z}%
^{q}}\sum_{s=0}^{S}\frac{C_{s}(\mathbf{k}+e\mathbf{A})_{l}}{\sqrt{(\mathbf{k}%
+e\mathbf{A})^{2}+M_{s}^{2}}},  \label{jlreg}
\end{eqnarray}%
with $C_{0}=1$ and $M_{0}=m$. Under the conditions $%
\sum_{s=0}^{S}C_{s}M_{s}^{2n}=0$,$\;n=0,1,\ldots ,[D/2]$, these expressions
are finite. After the renormalization subtractions the regulator is removed
taking the limit $M_{s}\rightarrow \infty $, $s\geqslant 1$. From formula (%
\ref{j0reg}) it follows that the VEV of the temporal component of the
fermionic current is renormalized to zero. Shifting the integration variable
in (\ref{jlreg}), we directly see that $\langle 0|j^{l}|0\rangle _{\text{reg}%
}=0$ for the components with $l=1,\ldots ,p$. Hence, the renormalized VEV of
the fermionic current is different from zero only for the components along
the compact dimensions.

Shifting the integration variables, $k_{l}+eA_{l}\rightarrow k_{l}$, $%
l=1,\ldots ,p$, for these components one finds%
\begin{equation}
\langle 0|j^{r}|0\rangle _{\text{reg}}=\frac{N}{2V_{q}}\int \frac{d\mathbf{k}%
_{p}}{(2\pi )^{p}}\sum_{\mathbf{n}_{q}\in \mathbf{Z}^{q}}\sum_{s=0}^{S}\frac{%
2\pi C_{s}(n_{r}+\tilde{\alpha}_{r})/L_{r}}{\sqrt{\mathbf{k}%
_{p}^{2}+\sum_{l=p+1}^{D}[2\pi (n_{l}+\tilde{\alpha}%
_{l})/L_{l}]^{2}+M_{s}^{2}}},  \label{jr}
\end{equation}%
with $r=p+1,\ldots ,D$, and we have introduced the notation%
\begin{equation}
\tilde{\alpha}_{l}=\alpha _{l}+eA_{l}L_{l}/(2\pi ).  \label{alfatilde}
\end{equation}%
Hence, the VEV of the fermionic current depends on components of the vector
potential along the compact dimensions alone. We present the second term on
the right of (\ref{alfatilde}) as $eA_{l}L_{l}/(2\pi )=N_{l}+\gamma _{l}$,
where $N_{l}$ is an integer number and $\gamma _{l}$ is the fractional part.
As it is seen from formula (\ref{jr}), only the fractional part leads to
nontrivial effects. Another point to be mentioned is that the presence of a
gauge field leads to the shift of the phases in the quasiperiodic boundary
conditions along compact dimensions. This feature is applicable to the
fermionic condensate and the VEV of the energy-momentum tensor as well. In
particular, the formulae for these quantities in the presence of a constant
gauge field are obtained from the corresponding formulae in Ref. \cite%
{Bell09a} by the replacement $\alpha _{l}\rightarrow \tilde{\alpha}_{l}$.

The property that the VEVs depend on the phases $\alpha _{l}$ and on the
vector potential components along compact dimensions in the combination (\ref%
{alfatilde}) can also be seen by the gauge transformation $A_{\mu }=A_{\mu
}^{\prime }+\partial _{\mu }\Lambda (x)$, $\psi (x)=\psi ^{\prime
}(x)e^{-ie\Lambda (x)}$, with the function $\Lambda (x)=A_{\mu }x^{\mu }$.
The new function $\psi ^{\prime }(x)$ satisfies Dirac equation with $A_{\mu
}^{\prime }=0$ and the quasiperiodicity conditions similar to (\ref{BC1})
with the replacement $\alpha _{l}\rightarrow \tilde{\alpha}_{l}$.
Corresponding eigenspinors are given by expressions (\ref{psibet+}), (\ref%
{psibet-}) with $\mathbf{A}=0$ and the eigenvalues for the wave
vector components along compact dimensions are defined by
$k_{l}=2\pi (n_{l}+\tilde{\alpha}_{l})/L_{l}$. In the new gauge,
the regularized VEVs are given by Eqs. (\ref{j0reg}) and
(\ref{jlreg}) with $\mathbf{A}=0$. The latter coincide with
(\ref{j0reg}) and (\ref{jlreg}) after the shift
$k_{l}+eA_{l}\rightarrow k_{l}$, $l=1,\ldots ,p$, of the
integration variables for the components along uncompactified
dimensions.

We will evaluate the VEV of fermionic current by two equivalent
methods: by applying the Abel-Plana type summation formula and
using the
zeta-function technique. In the first approach we apply to the series over $%
n_{r}$ in Eq. (\ref{jr}) the following summation formula:
\begin{eqnarray}
&&\sum_{n_{r}=-\infty }^{\infty }g(n_{r}+\tilde{\alpha}_{r})f(|n_{r}+\tilde{%
\alpha}_{r}|)=\int_{0}^{\infty }du\,\left[ g(u)+g(-u)\right] f(u)  \notag \\
&&\quad +i\int_{0}^{\infty }du\left[ f(iu)-f(-iu)\right] \sum_{\lambda =\pm
1}\frac{g(i\lambda u)}{e^{2\pi (u+i\lambda \tilde{\alpha}_{r})}-1}.
\label{sumform}
\end{eqnarray}%
This formula is obtained by combining the summation formulae given in Ref.
\cite{Beze08} (see also \cite{Bell09a}). In the special case of $g(x)=1$, $%
\tilde{\alpha}_{r}=0$ formula (\ref{sumform}) reduces to the standard
Abel-Plana formula (for the applications of the Abel-Plana formula and its
generalizations in quantum field theory see \cite{Most97,Mama76,Saha08b}).
Taking in Eq. (\ref{sumform})
\begin{equation}
g(x)=2\pi x/L_{r},\;f(x)=\left[ \mathbf{k}_{p}^{2}+\omega _{\mathbf{n}%
_{q-1}^{r}}^{2}+(2\pi x/L_{r})^{2}\right] ^{-1/2},\;  \label{gf}
\end{equation}%
with%
\begin{equation}
\omega _{\mathbf{n}_{q-1}^{r}}^{2}=\sum_{l=p+1,l\neq r}^{D}[2\pi (n_{l}+%
\tilde{\alpha}_{l})/L_{l}]^{2}+m^{2},  \label{omnq}
\end{equation}%
and $\mathbf{n}_{q-1}^{r}=(n_{p+1},\ldots ,n_{r-1},n_{r+1},\ldots
,n_{D})$, we see that the first integral on the right-hand side of
this formula vanishes. The contribution of the second integral to
the regularized VEV is finite in the limit $M_{s}\rightarrow
\infty $, $s\geqslant 1$, and in this term the
regulator can be safely removed. By using the expansion $1/(e^{y}-1)=%
\sum_{n=1}^{\infty }e^{-ny}$ in the integrand of the second integral, the
integrals with the separate terms in this expansion are evaluated explicitly
and one finds
\begin{eqnarray}
&&\sum_{n_{r}=-\infty }^{+\infty }\frac{2\pi (n_{r}+\tilde{\alpha}_{r})/L_{r}%
}{\sqrt{\mathbf{k}_{p}^{2}+\sum_{l=p+1}^{D}[2\pi (n_{l}+\tilde{\alpha}%
_{l})/L_{l}]^{2}+m^{2}}}  \notag \\
&&\quad =\frac{2L_{r}}{\pi }\sum_{n=1}^{\infty }\sin (2\pi n\tilde{\alpha}%
_{r})\sqrt{\mathbf{k}_{p}^{2}+\omega _{\mathbf{n}_{q-1}^{r}}^{2}}K_{1}(nL_{r}%
\sqrt{\mathbf{k}_{p}^{2}+\omega _{\mathbf{n}_{q-1}^{r}}^{2}}),
\label{nrseries}
\end{eqnarray}%
where $K_{\nu }(x)$ is the modified Bessel function. The equality in Eq. (%
\ref{nrseries}) is understood in the sense of the renormalized value.

By taking into account the result (\ref{nrseries}), from Eq. (\ref{jr}),
after the integration over $\mathbf{k}_{p}$, for the renormalized VEV one
finds%
\begin{equation}
\langle 0|j^{r}|0\rangle =\frac{2NL_{r}}{(2\pi )^{p/2+1}V_{q}}%
\sum_{n=1}^{\infty }\frac{\sin (2\pi n\tilde{\alpha}_{r})}{(nL_{r})^{p/2}}%
\sum_{\mathbf{n}_{q-1}^{r}\in \mathbf{Z}^{q-1}}\omega _{\mathbf{n}%
_{q-1}^{r}}^{p/2+1}K_{p/2+1}(nL_{r}\omega _{\mathbf{n}_{q-1}^{r}}),
\label{jr1}
\end{equation}%
with $\omega _{\mathbf{n}_{q-1}^{r}}$ defined by Eq. (\ref{omnq}).
As it is seen from this formula, the VEV of fermionic current is a
periodic function of $A_{l}L_{l}$ with the period of the flux
quantum $\Phi _{0}=2\pi
/e$ ($2\pi \hbar c/e$ in standard units). It is antisymmetric about $\tilde{%
\alpha}_{r}=1/2$. In the absence of the gauge field the VEV of
fermionic current vanishes for special cases of untwisted and
twisted fields. Of course, this result directly follows from the
symmetry of the problem for these special cases under the
reflection $z^{r}\rightarrow -z^{r} $. Note that $L_{r}^{D}\langle
0|j^{r}|0\rangle $ is a function of the
ratios $L_{l}/L_{r}$ and $mL_{r}$. As expected, in the large mass limit, $%
mL_{r}\gg 1$, the fermionic current along the direction $z^{r}$ is
exponentially suppressed.

Let us consider asymptotic limits of the VEV of fermionic current. For large
values of $L_{r}$, $L_{r}/L_{l}\gg 1$, the main contribution comes from the
term with $n=1$ and to the leading order we have%
\begin{equation}
\langle 0|j^{r}|0\rangle =\frac{NL_{r}\sin (2\pi \tilde{\alpha}_{r})}{(2\pi
L_{r})^{(p+1)/2}V_{q}}\omega _{0}^{(p+1)/2}e^{-L_{r}\omega _{0}},
\label{largeLr}
\end{equation}%
where $\omega _{0}^{2}=\sum_{l=p+1,l\neq r}^{D}(2\pi \beta
_{l}/L_{l})^{2}+m^{2}$ and $\beta _{l}=\min (|n_{l}+\tilde{\alpha}_{l}|)$.

In the limit when the length of the one of the compactified dimensions, say $%
z^{l}$, $l\neq r$, is large, $L_{l}\rightarrow \infty $, the main
contribution into the sum over $n_{l}$ in Eq. (\ref{jr1}) comes from large
values of $|n_{l}|$ and we can replace the summation by the integration in
accordance with%
\begin{equation}
\frac{1}{L_{l}}\sum_{n_{l}=-\infty }^{+\infty }f(2\pi |n_{l}+\tilde{\alpha}%
_{l}|/L_{l})=\frac{1}{\pi }\int_{0}^{\infty }dy\,f(y).  \label{sumtoint}
\end{equation}%
The integral over $y$ is evaluated by using the formula from Ref. \cite%
{Prud86} and from Eq. (\ref{jr1}) the corresponding formula is obtained for
the topology $R^{p+1}\times (S^{1})^{q-1}$.

Now let us consider the limit when the length of one of the compact
dimensions, say $z^{D}$, is small compared with $L_{r}$: $L_{D}\ll L_{r}$.
In this case, in the summation over $n_{D}$ the main contribution comes from
the term with minimum value of $|n_{D}+\tilde{\alpha}_{D}|$. If the
parameter $\tilde{\alpha}_{D}$ is an integer, the dominant contribution
comes from the term with $n_{D}=-\tilde{\alpha}_{D}$ (zero mode along the
direction $z^{D}$) and from (\ref{jr1}) we obtain:%
\begin{equation}
\langle 0|j^{r}|0\rangle =\frac{\delta _{D}}{L_{D}}\langle 0|j^{r}|0\rangle
_{R^{p}\times (S^{1})^{D-1-p}},  \label{jrsmallLD}
\end{equation}%
where $\langle 0|j^{r}|0\rangle _{R^{p}\times (S^{1})^{D-1-p}}$ is the
fermionic current in $(D-1)$-dimensional space with spatial topology $%
R^{p}\times (S^{1})^{D-1-p}$ and with the lengths of compact dimensions $%
L_{p+1},\ldots ,L_{D-1}$. In Eq. (\ref{jrsmallLD}), $\delta _{D}=1$ for even
$D$ and $\delta _{D}=2$ for odd $D$. If the parameter $\tilde{\alpha}_{D}$
is non-integer and $\beta _{D}L_{r}/L_{D}\gg 1$, the argument of the
modified Bessel function in Eq. (\ref{jr1}) is large. By using the
corresponding asymptotic formula, to the leading order we find
\begin{equation}
\langle 0|j^{r}|0\rangle \approx \frac{4NL_{r}^{-p}}{(2\pi )^{(p+3)/2}V_{q}}%
\sum_{n=1}^{\infty }\frac{\sin (2\pi n\tilde{\alpha}_{r})}{n^{(p+1)/2}}\sum_{%
\mathbf{n}_{q-2}^{r}\in \mathbf{Z}^{q-2}}b_{\mathbf{n}%
_{q-2}^{r}}^{(p+1)/2}e^{-nb_{\mathbf{n}_{q-2}^{r}}},  \label{jrsmallLDb}
\end{equation}%
where
\begin{equation}
b_{\mathbf{n}_{q-2}^{r}}^{2}=\left( 2\pi \beta _{D}L_{r}/L_{D}\right)
^{2}+\sum_{l=p+1,l\neq r}^{D-1}[2\pi (n_{l}+\tilde{\alpha}%
_{l})L_{r}/L_{l}]^{2}+L_{r}^{2}m^{2},  \label{bnq-2}
\end{equation}%
with $\mathbf{n}_{q-2}^{r}=(n_{p+1},\ldots ,n_{r-1},n_{r+1},\ldots ,n_{D-1})$
. In this case the VEV of fermionic current is exponentially suppressed.

\begin{figure}[tbph]
\begin{center}
\epsfig{figure=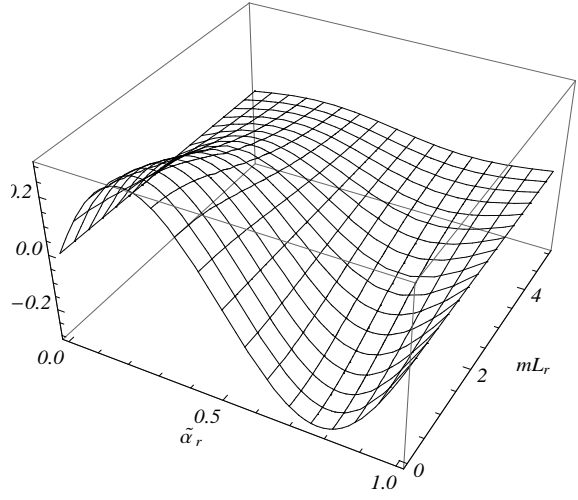,width=7.5cm,height=6.5cm}
\end{center}
\caption{The VEV of fermionic current, $L_{r}^{4}\langle 0|j^{r}|0\rangle $,
in the model with spatial topology $R^{3}\times S^{1}$ as a function of the
phase $\tilde{\protect\alpha}_{r}$ and the parameter $mL_{r}$.}
\label{fig1}
\end{figure}

In the special case with a single compact dimension we have $p=D-1$, $q=1$, $%
\omega _{\mathbf{n}_{q-1}^{r}}=m$, and the general formula (\ref{jr1})
simplifies to%
\begin{equation}
\langle 0|j^{r}|0\rangle =\frac{2Nm^{(D+1)/2}}{(2\pi )^{(D+1)/2}}%
\sum_{n=1}^{\infty }\sin (2\pi n\tilde{\alpha}_{r})\frac{K_{(D+1)/2}(nL_{r}m)%
}{(nL_{r})^{(D-1)/2}}.  \label{jr1q1}
\end{equation}%
For a massless field this expression takes the form%
\begin{equation}
\langle 0|j^{r}|0\rangle =\frac{N\Gamma ((D+1)/2)}{\pi ^{(D+1)/2}L_{r}^{D}}%
\sum_{n=1}^{\infty }\frac{\sin (2\pi n\tilde{\alpha}_{r})}{n^{D}}.
\label{jrm0}
\end{equation}%
For odd values $D$ the series in this formula is summed in terms of the
Bernoulli polynomials $B_{D}(x)$ and one finds%
\begin{equation}
\langle 0|j^{r}|0\rangle =(-1)^{(D+1)/2}\frac{2^{(D-1)/2}\pi ^{D/2}}{\Gamma
(D/2+1)L_{r}^{D}}B_{D}(\tilde{\alpha}_{r}),  \label{jrm0odd}
\end{equation}%
for $0\leqslant \tilde{\alpha}_{r}\leqslant 1$. In figure \ref{fig1} we
plot the VEV of fermionic current in the simplest Kaluza-Klein model with $%
D=4$ as a function of parameters $\tilde{\alpha}_{r}$ and $mL_{r}$. In
Kaluza-Klein type models the fermionic current with the components along
compact dimensions is a source of cosmological magnetic fields.

\section{Zeta function approach}

\label{sec:FCZeta}

In this section, for the evaluation of the VEV of fermionic current we
follow a different route based on the zeta function method \cite%
{Eliz94,Eliz95,Kirs01}. This allows us to obtain an alternative
representation. To start we note that the mode sum for the fermionic current
can be written as%
\begin{equation}
\langle 0|j^{r}|0\rangle =\frac{\pi N}{L_{r}^{2}}\sum_{n_{r}=-\infty
}^{+\infty }(n_{r}+\tilde{\alpha}_{r})\zeta (1/2),  \label{jrzeta}
\end{equation}%
where the generalized zeta function%
\begin{equation}
\zeta (s)=\frac{L_{r}}{V_{q}}\int \frac{d\mathbf{k}_{p}}{(2\pi )^{p}}\sum_{%
\mathbf{n}_{q-1}^{r}\in \mathbf{Z}^{q-1}}\{\mathbf{k}_{p}^{2}+\sum_{l=p+1,%
\neq r}^{D}[2\pi (n_{l}+\tilde{\alpha}_{l})/L_{l}]^{2}+m_{r}^{2}\}^{-s},
\label{zeta}
\end{equation}%
is introduced with the notation
\begin{equation}
m_{r}^{2}=m^{2}+[2\pi (n_{r}+\tilde{\alpha}_{r})/L_{r}]^{2}.  \label{mr2}
\end{equation}%
As it follows from Eq. (\ref{jrzeta}), for the evaluation of the
renormalized VEV of fermionic current we need to have the analytic
continuation of the zeta function $\zeta (s)$ at the point $s=1/2$.

With this aim we first integrate over the wave vector along the
uncompactified dimensions:%
\begin{equation}
\zeta (s)=\frac{\Gamma (s-p/2)L_{r}}{(4\pi )^{p/2}\Gamma (s)V_{q}}\sum_{%
\mathbf{n}_{q-1}^{r}\in \mathbf{Z}^{q-1}}\left\{ \sum_{l=p+1,\neq
r}^{D}[2\pi (n_{l}+\tilde{\alpha }_{l})/L_{l}]^{2}+m_{r}^{2}\right\}
^{p/2-s}.  \label{zeta1}
\end{equation}%
An exponentially convergent expression for the analytic continuation of the
multiseries in Eq. (\ref{zeta1}) can be obtained by using the generalized
Chowla-Selberg formula \cite{Eliz98}. The application of this formula to Eq.
(\ref{zeta1}) gives the following result%
\begin{eqnarray}
\zeta (s) &=&\frac{m_{r}^{D-2s-1}}{(4\pi )^{(D-1)/2}}\frac{\Gamma (s-(D-1)/2)%
}{\Gamma (s)}+\frac{2^{1-s}m_{r}^{D-2s-1}}{(2\pi )^{(D-1)/2}\Gamma (s)}
\notag \\
&&\times \sideset{}{'}{\sum}_{\mathbf{n}_{q-1}^{r}\in \mathbf{Z}^{q-1}}\cos
(2\pi \mathbf{n}_{q-1}^{r}\cdot \boldsymbol{\alpha }%
_{q-1})f_{(D-1)/2-s}(m_{r}g(\mathbf{L}_{q-1},\mathbf{n}_{q-1}^{r})),
\label{zetadec}
\end{eqnarray}%
with $\mathbf{L}_{q-1}=(L_{p+1},\ldots ,L_{r-1},L_{r+1},\ldots L_{D})$ and $%
\boldsymbol{\alpha }_{q-1}=(\tilde{\alpha }_{p+1},\ldots ,\tilde{\alpha }%
_{r-1},\tilde{\alpha }_{r+1},\ldots \tilde{\alpha }_{D})$. In Eq. (\ref%
{zetadec}) we have introduced the notation $f_{\nu }(x)=K_{\nu }(x)/x^{\nu }$%
. The prime on the summation sign in (\ref{zetadec}) means that the term $%
\mathbf{n}_{q}=0$ should be excluded from the sum and
\begin{equation}
g(\mathbf{L}_{q-1},\mathbf{n}_{q-1}^{r})=\Big(\sum_{i=p+1,\neq
r}^{D}L_{i}^{2}n_{i}^{2}\Big)^{1/2}.  \label{gLm}
\end{equation}

The part in the fermionic current containing the second term on
the right-hand side of Eq. (\ref{zetadec}) is finite at the
physical point and, hence, the analytic continuation is needed for
the part with the first term alone. In order to do this, we apply
the summation formula (\ref{sumform}) to the corresponding series
over $n_{r}$. After transformations similar to those already used
in the derivation of Eq. (\ref{jr1}) and by making use of the
standard properties of the gamma function, one finds%
\begin{eqnarray}
&&\frac{\Gamma (s-(D-1)/2)}{2L_{r}(4\pi )^{(D-1)/2}\Gamma (s)}%
\sum_{n_{r}=-\infty }^{+\infty }\frac{2\pi (n_{r}+\tilde{\alpha}_{r})/L_{r}}{%
m_{r}^{2s+1-D}}  \notag \\
&&\quad =\frac{(2m)^{(D-2s)/2+1}}{(4\pi )^{D/2}\Gamma (s)}\sum_{n=1}^{\infty
}\sin (2\pi n\tilde{\alpha}_{r})\frac{K_{(D-2s)/2+1}(nL_{r}m)}{%
(nL_{r})^{(D-2s)/2}}.  \label{Rel2}
\end{eqnarray}%
As it can be easily checked, the right-hand side of this relation is finite
at $s=1/2$. Combining Eqs. (\ref{jrzeta}), (\ref{zetadec}), (\ref{Rel2}),
for the VEV of fermionic current we find the following representation%
\begin{eqnarray}
\langle 0|j^{r}|0\rangle  &=&\frac{2Nm^{(D+1)/2}}{(2\pi )^{(D+1)/2}}%
\sum_{n=1}^{\infty }\sin (2\pi n\tilde{\alpha}_{r})\frac{K_{(D+1)/2}(nL_{r}m)%
}{(nL_{r})^{(D-1)/2}}+\frac{NL_{r}^{-2}}{(2\pi )^{D/2-1}}\sum_{n_{r}=-\infty
}^{+\infty }(n_{r}+\tilde{\alpha}_{r})  \notag \\
&&\times m_{r}^{D-2}\sideset{}{'}{\sum}_{\mathbf{n}_{q-1}^{r}\in \mathbf{Z}%
^{q-1}}\cos (2\pi \mathbf{n}_{q-1}^{r}\cdot \boldsymbol{\alpha }%
_{q-1})f_{D/2-1}(m_{r}g(\mathbf{L}_{q-1},\mathbf{n}_{q-1}^{r})).  \label{jr2}
\end{eqnarray}%
Note that in the limit $L_{l}\rightarrow \infty $, $l\neq r$, the second
term on the right-hand side of this formula vanishes and we obtain the
vacuum fermionic current in the model with a single compact dimension. The
latter coincides with Eq. (\ref{jr1q1}).

Formula (\ref{jr2}) is further simplified by using the relation%
\begin{eqnarray}
&&\sum_{n_{r}=-\infty }^{+\infty }\left[ 2\pi (n_{r}+\tilde{\alpha}%
_{r})/L_{r}\right] m_{r}^{D-2}f_{D/2-1}(m_{r}y)  \notag \\
&&\quad =\sqrt{\frac{2}{\pi }}L_{r}^{2}m^{D+1}\sum_{n_{r}=1}^{+\infty
}n_{r}\sin (2\pi n_{r}\tilde{\alpha}_{r})f_{(D+1)/2}(m\sqrt{%
y^{2}+n_{r}^{2}L_{r}^{2}}).  \label{Rel3}
\end{eqnarray}%
This relation is obtained by integrating the Poisson's resummation formula $%
\sum_{n=-\infty }^{+\infty }F(x)\delta (x-n)=\sum_{n=-\infty
}^{+\infty }F(x)e^{2i\pi nx}$ with the function $F(x)$ defined by
the left-hand side of Eq. (\ref{Rel3}). The integral for the
right-hand side is evaluated using the formula from Ref.
\cite{Prud86}. By taking into account Eq. (\ref{Rel3}), from Eq.
(\ref{jr2}) we find
\begin{eqnarray}
\langle 0|j^{r}|0\rangle  &=&\frac{2Nm^{D+1}L_{r}}{(2\pi )^{(D+1)/2}}%
\sum_{n_{r}=1}^{\infty }n_{r}\sin (2\pi n_{r}\tilde{\alpha}_{r})  \notag \\
&&\times \sum_{\mathbf{n}_{q-1}^{r}\in \mathbf{Z}^{q-1}}\cos (2\pi \mathbf{n}%
_{q-1}^{r}\cdot \boldsymbol{\alpha }_{q-1})f_{(D+1)/2}(mg(\mathbf{L}_{q},%
\mathbf{n}_{q})),  \label{jr3}
\end{eqnarray}%
with the notation%
\begin{equation}
g(\mathbf{L}_{q},\mathbf{n}_{q})=\Big(\sum_{i=p+1}^{D}L_{i}^{2}n_{i}^{2}\Big)%
^{1/2}.  \label{glqng}
\end{equation}%
The equivalence of two representations (\ref{jr1}) and (\ref{jr3})
for the VEV\ of fermionic current is seen using the relation
\begin{eqnarray}
&&\sum_{\mathbf{n}_{q-1}^{r}\in \mathbf{Z}^{q-1}}\cos (2\pi \mathbf{n}%
_{q-1}^{r}\cdot \boldsymbol{\alpha }_{q-1})f_{(D+1)/2}(mg(\mathbf{L}_{q},%
\mathbf{n}_{q}))  \notag \\
&&\quad =\frac{(2\pi )^{(q-1)/2}L_{r}}{V_{q}m^{D+1}}\sum_{\mathbf{n}%
_{q-1}^{r}\in \mathbf{\ Z}^{q-1}}\omega _{\mathbf{n}%
_{q-1}^{r}}^{(D-q)/2+1}f_{(D-q)/2+1}(nL_{r}\omega _{\mathbf{n}_{q-1}^{r}}).
\label{Rel4}
\end{eqnarray}%
The proof of this relation can be found in Appendix of Ref. \cite{Bell09a}.
The advantage of the representation (\ref{jr1}), as compared with Eq. (\ref%
{jr3}), is that in the case of a massless field, for large values of $n_{l}$
the separate terms in the multiseries decay exponentially instead of
power-law decay in Eq. (\ref{jr3}).

\section{Fermionic current in carbon nanotubes}

\label{sec:FCNano}

Carbon nanotubes have attracted much attention recently due to the
experimental observation of a number of novel electronic
properties. In this section we apply general results obtained
above for the electrons in cylindrical and toroidal carbon
nanotubes. A single-wall cylindrical nanotube is a graphene sheet
rolled into a cylindrical shape. The electronic band structure of
graphene close to the Dirac points shows a conical dispersion
$E(\mathbf{k})=v_{F}|\mathbf{k}|$, where $\mathbf{k}$ is the
momentum measured relatively to the Dirac points and $v_{F}\approx
10^{8}$ cm/s represents the Fermi velocity which plays the role of
a speed of light. The corresponding low-energy excitations can be
described by a pair of two-component spinors, $\psi _{A}$ and
$\psi _{B}$, corresponding to the two different triangular
sublattices of the honeycomb lattice of graphene (see, for
instance, \cite{Sait98,Seme84}). The Dirac equation for these
spinors
has the form%
\begin{equation}
(iv_{F}^{-1}\gamma ^{0}D_{0}+i\gamma ^{l}D_{l}-m)\psi _{J}=0,\
\label{DeqGraph}
\end{equation}%
where $J=A,B$, $l=1,2$, and $D_{\mu }$ is defined in Eq. (\ref{Direq}) with $%
e=-|e|$ for electrons. To keep the discussion general we have included in
Eq. (\ref{DeqGraph}) the mass (gap) term. The gap in the energy spectrum is
essential in many physical application. This gap can be generated by a
number of mechanisms (see, for example, \cite{Seme84,Cham00,Giov07,Seme08}).
In particular, they include the breaking of symmetry between two sublattices
by introducing a staggered onsite energy \cite{Seme84} and the deformations
of bonds in the graphene lattice \cite{Cham00}. Another approach is to
attach a graphene monolayer to a substrate the interaction with which breaks
the sublattice symmetry \cite{Giov07}. For metallic nanotubes we have
periodic boundary conditions ($\alpha _{l}=0$) along the compact dimension
and for semiconductor nanotubes, depending on the chiral vector, we have two
classes of inequivalent boundary conditions corresponding to $\alpha
_{l}=\pm 1/3$. These phases have opposite signs for the sublattices $A$ and $%
B$.

The presence of the gauge field in Eq. (\ref{DeqGraph}) leads to the
Aharonov-Bohm effect in carbon nanotubes \cite{AhBohm}. This effect
manifests itself in a periodic energy gap modulation and conductance
oscillations as a function of enclosed magnetic flux with a period of the
order of the flux quantum. Similar oscillations arise in the VEV\ of
fermionic current along compact dimensions. We consider the cases of
cylindrical and toroidal nanotubes separately.

\subsection{The case $D=1$}

\begin{figure}[tbph]
\begin{center}
\begin{tabular}{cc}
\epsfig{figure=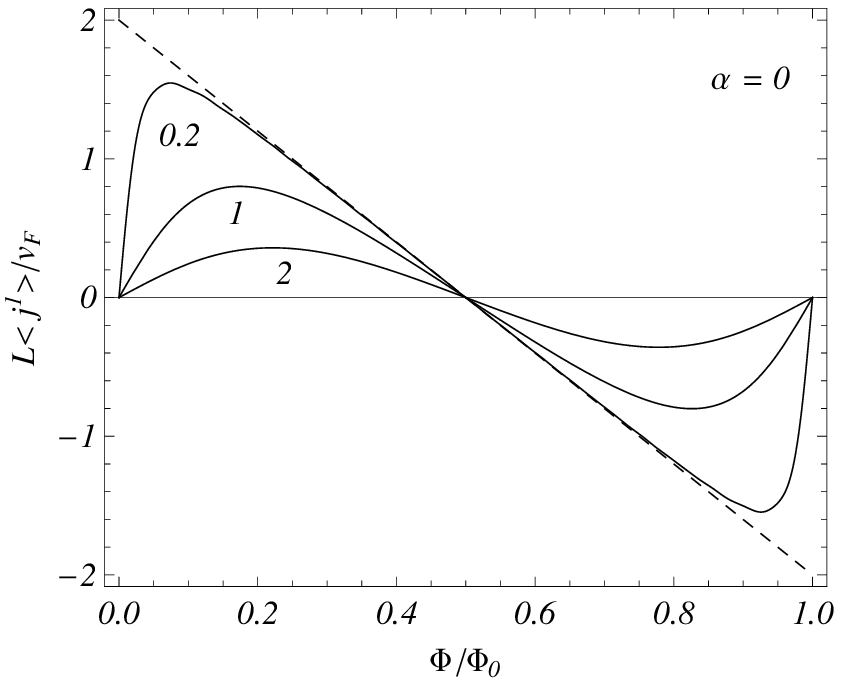,width=7.cm,height=6.cm} & \quad %
\epsfig{figure=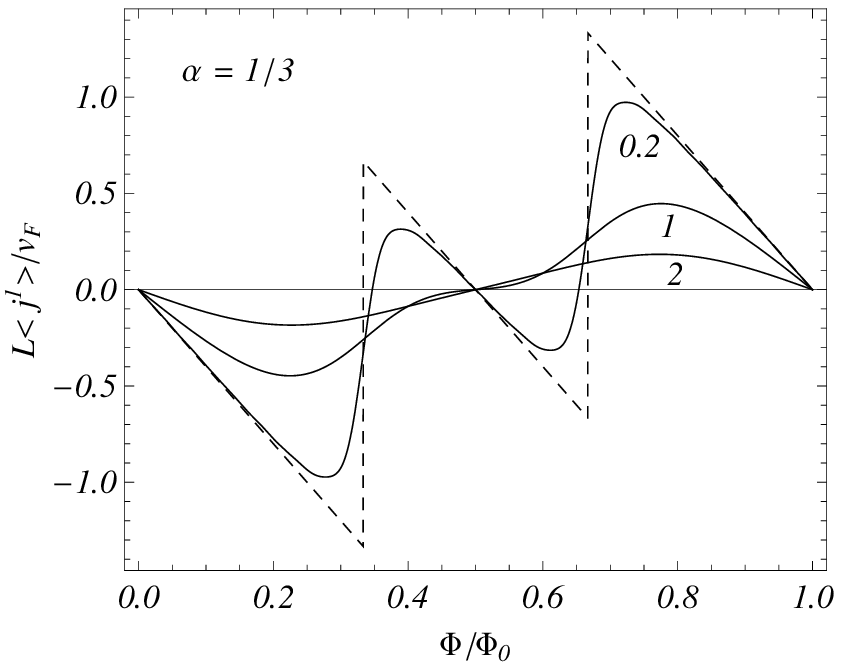,width=7.cm,height=6.cm}%
\end{tabular}%
\end{center}
\caption{The VEV of fermionic current in $D=1$ model for the
periodicity conditions with $\protect\alpha =0$ (left panel) and
$\protect\alpha =1/3$ (right panel) as a function of the magnetic
flux.} \label{fig2}
\end{figure}

We start with the simplest case $D=1$ with a compact dimension of the length
$L_{1}=L$. The corresponding phase in the periodicity condition we denote $%
\alpha _{1}=\alpha $. As it is seen below, this case can be considered as a
model of a toroidal nanotube in the limit when the length of the one of
compact dimensions is small. The corresponding effective two-dimensional
Dirac-like theory is discussed in Refs. \cite{Sasa02}. By summing the
contributions coming from two sublattices with opposite signs of $\alpha $,
for the VEV of fermionic current one finds%
\begin{equation}
\langle j^{1}\rangle =\frac{4v_{F}m}{\pi }\sum_{n=1}^{\infty }\cos (2\pi
n\alpha )\sin (2\pi n\Phi /\Phi _{0})K_{1}(mLn),  \label{jrD1}
\end{equation}%
where $2\pi \Phi /\Phi _{0}=eA_{1}L/(\hbar c)$ with $\Phi $ being the
magnetic flux. The corresponding vector potential can be generated by the
magnetic field perpendicular to the plane of torus and located inside a
coaxial cylinder with radius smaller than $L/(2\pi )$. For a massless case
from here we have $\langle j^{1}\rangle =v_{F}\sum_{j=\pm }\mathcal{I}(\Phi
/\Phi _{0}+j\alpha )/L$, where we have defined the function%
\begin{equation}
\mathcal{I}(x)=\frac{2}{\pi }\sum_{n=1}^{\infty }\frac{\sin (2\pi nx)}{n}%
=\left\{
\begin{array}{cc}
1-2\{x\}, & \{x\}>0 \\
2|\{x\}|-1, & \{x\}<0%
\end{array}%
\right. .  \label{Ix}
\end{equation}%
The fractional part $\{x\}$ on the right-hand side of this formula is
defined in accordance with the Mathematica function FractionalPart[$x$]. In
figure \ref{fig2} we have plotted the VEV (\ref{jrD1}) as a function of the
magnetic flux for different values of the parameter $mL$ (numbers near the
curves). The dashed lines correspond to a massless case. For the left and
right panels $\alpha =0$ and $\alpha =1/3$, respectively. The electric
current corresponding to the VEV of the fermionic current is of order $%
|e|v_{F}/L$. Note that the persistent currents in normal metal rings with
this order of magnitude have been recently measured in Refs. \cite{Bluh09}.

\subsection{Cylindrical nanotubes}

A single-wall cylindrical nanotube is a rolled-up graphene sheet in the
hollow cylindrical structure. For the case of cylindrical nanotube we have
spatial topology $R^{1}\times S^{1}$ with the compactified dimension of
length $L$. The nanotube is characterized by its chiral vector $\mathbf{C}%
_{h}=(n_{w},m_{w})$, with $n_{w}$, $m_{w}$ being integers determining the
circumference in accordance with $L=|\mathbf{C}_{h}|=a\sqrt{%
n_{w}^{2}+m_{w}^{2}+n_{w}m_{w}}$. Here $a=2.46\mathring{A}$ is the lattice
constant for graphene. A zigzag nanotube corresponds to the special case $%
\mathbf{C}_{h}=(n_{w},0)$, and a armchair nanotube corresponds to the case $%
\mathbf{C}_{h}=(n_{w},n_{w})$. All other cases correspond to chiral
nanotubes. The electronic properties of carbon nanotubes can be either
metallic or semiconductor-like depending on the chiral vector. In the case $%
n_{w}-m_{w}=3q_{w}$, $q_{w}\in Z$, the nanotube will be metallic and in the
case $n_{w}-m_{w}\neq 3q_{w}$ the nanotube will be semiconductor with an
energy gap inversely proportional to the diameter.

For the case under discussion $D=2$ and the general formula for the VEV of
fermionic current takes the form ($N=2$, $\alpha _{p+1}\equiv \alpha $)%
\begin{equation}
\langle 0|j^{2}|0\rangle =\frac{1}{\pi L^{2}}\sum_{n=1}^{\infty }\sin (2\pi n%
\tilde{\alpha})\frac{1+nLm}{n^{2}e^{nLm}},  \label{jrD2}
\end{equation}%
with the notation $\tilde{\alpha}=\alpha +eA_{2}L/(2\pi )$, and $\langle
0|j^{1}|0\rangle =0$. In metallic nanotubes $\alpha =0$ and for
semiconductor nanotubes $\alpha =\pm 1/3$. For graphene sheet we have two
spinors that describe Bloch states residing on the two different
sublattices. Summing the contributions from these sublattices and taking
into account that, for these two sublattices the phases $\alpha $ have
opposite signs, for the total fermionic current we find
\begin{equation}
\langle j^{2}\rangle ^{\text{(cn)}}=\frac{2v_{F}}{\pi L^{2}}%
\sum_{n=1}^{\infty }\cos (2\pi n\alpha )\sin (2\pi n\Phi /\Phi _{0})\frac{%
1+nLm}{n^{2}e^{nLm}},  \label{jrs}
\end{equation}%
where $\alpha =0,1/3$ for metallic and semiconductor nanotubes respectively.
In these formulae $2\pi \Phi /\Phi _{0}=eA_{2}L/(\hbar c)$ with $\Phi $
being the magnetic flux passing through the cross section of the nanotube.
Note that in Eq. (\ref{jrs}) (and in the formulae below), we give the
fermionic current for a given spin component. The total current is obtained
multiplying by the number of spin components which is 2 for graphene.

\begin{figure}[tbph]
\begin{center}
\begin{tabular}{cc}
\epsfig{figure=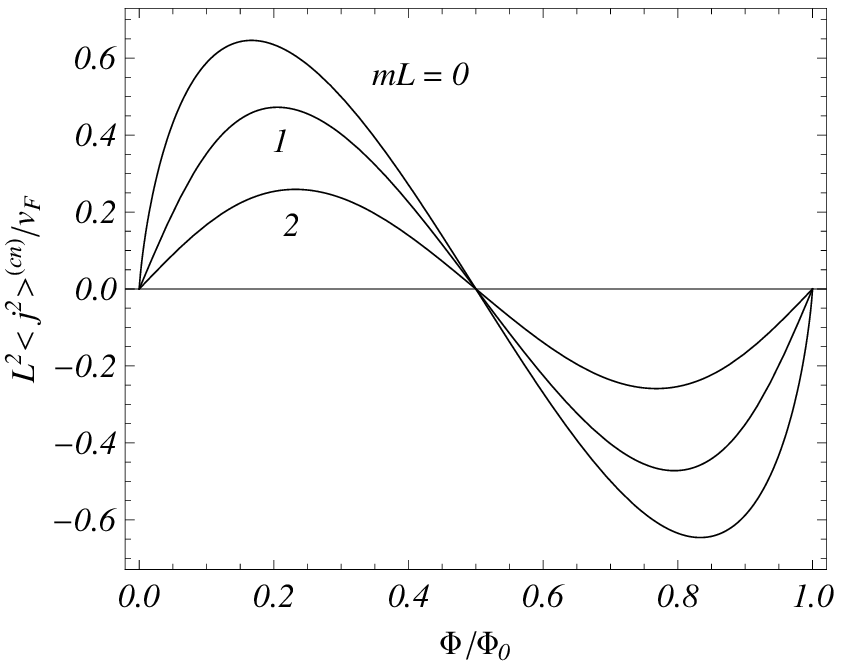,width=7.cm,height=6.cm} & \quad %
\epsfig{figure=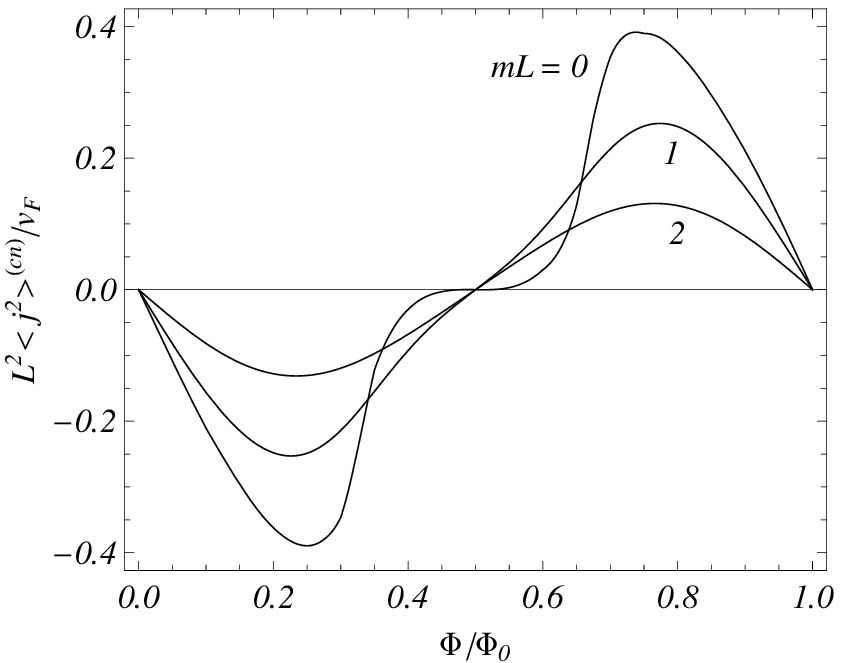,width=7.cm,height=6.cm}%
\end{tabular}%
\end{center}
\caption{The VEV of fermionic current in metallic (left panel) and
semiconductor (right panel) nanotubes as a function of the magnetic flux.}
\label{fig3n}
\end{figure}

As it is seen from Eq. (\ref{jrs}), in the absence of the magnetic flux the
total fermionic current vanishes due to the cancellation of contributions
from two sublattices. The magnetic flux breaks this symmetry and an
effective current appears. However, it should be noted that, in general, the
mass terms in the Dirac equation for separate sublattices can be different.
In this case an effective fermionic current appears without an external
magnetic field. In figure \ref{fig3n} we plot the VEV of the fermionic
current for various values of the parameter $mL$ (numbers near the curves)
in metallic (left panel) and semiconductor (right panel) cylindrical
nanotubes as a function of magnetic flux in units of magnetic flux quantum.

\subsection{Toroidal nanotubes}

A toroidal nanotube corresponds to a finite graphene sheet with the
periodical boundary conditions along the transverse and longitudinal
directions. This form of carbon structure was discovered in Refs. \cite%
{Liu97}. The carbon toroid is determined by its chiral, $\mathbf{C}%
_{h}=(n_{w},m_{w})$, and translational, $\mathbf{T}=(p_{w},q_{w})$, vectors.
The parameters $(n_{w},m_{w},p_{w},q_{w})$ define the geometric structure
and physical properties of toroidal nanotubes. For the geometry of a
toroidal nanotube we have the spatial topology $(S^{1})^{2}$ with $p=0$ and $%
q=2$ and the corresponding formulae for the VEV of fermionic current are
directly obtained from the general results (\ref{jr1}) and (\ref{jr3}) (on
the persistent currents in toroidal carbon nanotubes see Refs. \cite{Lin98}%
). For a graphene sheet we have two sublattices with opposite signs of the
phases $\alpha _{l}$. The total current is obtained by summing the
corresponding contributions and one finds%
\begin{eqnarray}
\langle j^{r}\rangle ^{\text{(tor)}} &=&\frac{2L_{r}v_{F}}{\pi }%
\sum_{n_{r}=1}^{\infty }n_{r}\sum_{n_{l}=-\infty }^{\infty }\frac{\cos [2\pi
\left( n_{1}\alpha _{1}+n_{2}\alpha _{2}\right) ]}{%
(L_{1}^{2}n_{1}^{2}+L_{2}^{2}n_{2}^{2})^{3/2}}  \notag \\
&&\times \frac{1+m\sqrt{L_{1}^{2}n_{1}^{2}+L_{2}^{2}n_{2}^{2}}}{\exp (m\sqrt{%
L_{1}^{2}n_{1}^{2}+L_{2}^{2}n_{2}^{2}})}\sin \left[ 2\pi \left( n_{1}\Phi
_{1}+n_{2}\Phi _{2}\right) /\Phi _{0}\right] ,  \label{jrtor1n}
\end{eqnarray}%
where $r,l=1,2,$ $l\neq r$, and $2\pi \Phi _{l}/\Phi _{0}=eA_{l}L_{l}/(\hbar
c)$. As in the case of cylindrical nanotubes, due to the cancellation of
contributions coming from separate sublattices, the fermionic current in
toroidal nanotubes vanishes in the absence of the magnetic flux. An
alternative representation is obtained by using formula (\ref{jr1}):%
\begin{eqnarray}
\langle j^{r}\rangle ^{\text{(tor)}} &=&\frac{2v_{F}}{\pi L_{l}^{2}}%
\sum_{\delta =\pm }\sum_{n_{r}=1}^{\infty }\sin (2\pi n_{r}\tilde{\alpha}%
_{r}^{(\delta )})\sum_{n_{l}=-\infty }^{+\infty }\sqrt{[2\pi (n_{l}+\tilde{%
\alpha}_{l}^{(\delta )})]^{2}+L_{l}^{2}m^{2}}  \notag \\
&&\times K_{1}(n_{r}(L_{r}/L_{l})\sqrt{[2\pi (n_{l}+\tilde{\alpha}%
_{l}^{(\delta )})]^{2}+L_{l}^{2}m^{2}}),  \label{jrtor2n}
\end{eqnarray}%
with $r,l=1,2,$ $l\neq r$, and $\tilde{\alpha}_{l}^{(\pm )}=\pm \alpha
_{l}+\Phi _{l}/\Phi _{0}$. This formula is further simplified when $A_{l}=0$:%
\begin{eqnarray}
\langle j^{r}\rangle ^{\text{(tor)}} &=&\frac{4v_{F}}{\pi L_{l}^{2}}%
\sum_{n=1}^{\infty }\sin (2\pi n_{r}\Phi _{r}/\Phi _{0})\cos (2\pi
n_{r}\alpha _{r})\sum_{n_{l}=-\infty }^{+\infty }\sqrt{[2\pi (n_{l}+\alpha
_{l})]^{2}+L_{l}^{2}m^{2}}  \notag \\
&&\times K_{1}(n_{r}(L_{r}/L_{l})\sqrt{[2\pi (n_{l}+\alpha
_{l})]^{2}+L_{l}^{2}m^{2}}).  \label{jrtor2b}
\end{eqnarray}%
Note that in this case the component $\langle 0|j^{l}|0\rangle ^{\text{(tor)}%
}$ is nonzero only for $\alpha _{1},\alpha _{2}\neq 0$.

Let us consider the asymptotic limit of the fermionic current in toroidal
nanotubes in the case $L_{1}\ll L_{2}$ for a fixed value of $mL_{2}$. First
we consider the component $\langle j^{1}\rangle ^{\text{(tor)}}$. For this
component in Eq. (\ref{jrtor2n}) one has $l=2$. In the limit under
consideration the main contribution to the series over $n_{2}$ comes from
large values and we can replace the summation by integration. The integral
is evaluated explicitly and to the leading order the expression for $\langle
j^{1}\rangle ^{\text{(tor)}}$ coincides with the corresponding result for
cylindrical nanotubes given by (\ref{jrs}) (with $L=L_{1}$). The behavior of
the component $\langle j^{2}\rangle ^{\text{(tor)}}$ crucially depends on
whether the parameter $\tilde{\alpha}_{1}^{(\delta )}$ is integer or not.
When this parameter is non-integer (for both $\delta =\pm $) the argument of
the modified Bessel function in Eq. (\ref{jrtor2b}) (with $r=2$, $l=1$) is
large and the dominant contribution comes from the term with $n_{2}=1$ and
from the term in the summation over $n_{1}$ with minimum value of $|n_{1}+%
\tilde{\alpha}_{1}^{(\delta )}|$. The VEV of fermionic current is
exponentially suppressed: $\langle j^{2}\rangle ^{\text{(tor)}}\sim \exp [-%
\sqrt{(2\pi \beta _{1}L_{2}/L_{1})^{2}+L_{2}^{2}m^{2}}]$, with $\beta
_{1}=\min |n_{1}+\tilde{\alpha}_{1}^{(\delta )}|$. For nanotubes metallic
along the direction $z^{1}$ with the length $L_{1}$ and for $A_{1}=0$ one
has $\tilde{\alpha}_{1}^{(\pm )}=0$. In this case and for $L_{1}\ll L_{2}$
the main contribution to $\langle j^{2}\rangle ^{\text{(tor)}}$ comes from
the term with $n_{1}=0$ in Eq. (\ref{jrtor2n}) and the quantity $%
L_{1}\langle j^{2}\rangle ^{\text{(tor)}}$ coincides with the corresponding
result in $D=1$ case (formula (\ref{jrD1}) with $L=L_{2}$ and $\alpha
=\alpha _{2}$). If $\alpha _{1}\neq 0$ and $\tilde{\alpha}_{1}^{(\delta )}$
is an integer (note that this can be satisfied for one of values $\delta $)
the main contribution comes from the term for which $n_{1}+\tilde{\alpha}%
_{1}^{(\delta )}=0$. In this case only one of the sublattices contributes to
the fermionic current.

\begin{figure}[tbph]
\begin{center}
\epsfig{figure=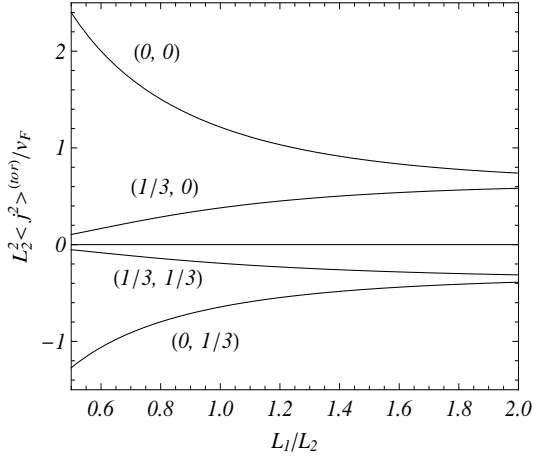,width=7.cm,height=6.cm}
\end{center}
\caption{The VEV of the current for a massless fermionic field in a toroidal
nanotube as a function of the ratio $L_{1}/L_{2}$ for different values of
the phases $(\protect\alpha _{1},\protect\alpha _{2})$ and for the magnetic
flux $\Phi _{2}/\Phi _{0}=0.2$, $\Phi _{1}=0$.}
\label{fig4n}
\end{figure}

In figure \ref{fig4n} we plot the dependence of the fermionic current $%
\langle j^{2}\rangle ^{\text{(tor)}}$ in the massless case as a function of
the ratio $L_{1}/L_{2}$ for different values of the phases $(\alpha
_{1},\alpha _{2})$ (numbers near the curves) and for $\Phi _{2}/\Phi _{0}=0.2
$, $\Phi _{1}=0$. In this case the component $\langle j^{1}\rangle ^{\text{%
(tor)}}$ is nonzero for $\alpha _{1,2}\neq 0$ only. As it was shown above
and clearly seen from figure, for large values of the ratio $L_{1}/L_{2}$
the component $\langle j^{2}\rangle ^{\text{(tor)}}$ of the fermionic
current tends to the corresponding quantity in a cylindrical nanotube with
circumference $L_{2}$. In the opposite limit of small values of $L_{1}/L_{2}$
the VEV\ tends to zero for semiconducting type periodicity condition along
the direction $z^{1}$. Again, this is in agreement with the asymptotic
analysis given before.

\begin{figure}[tbph]
\begin{center}
\begin{tabular}{cc}
\epsfig{figure=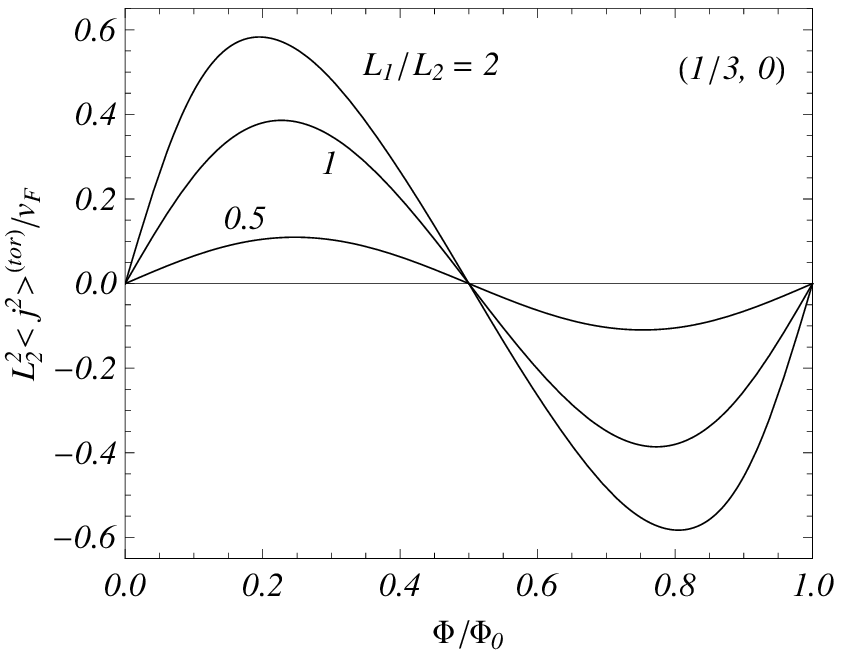,width=7.cm,height=6.cm} & \quad %
\epsfig{figure=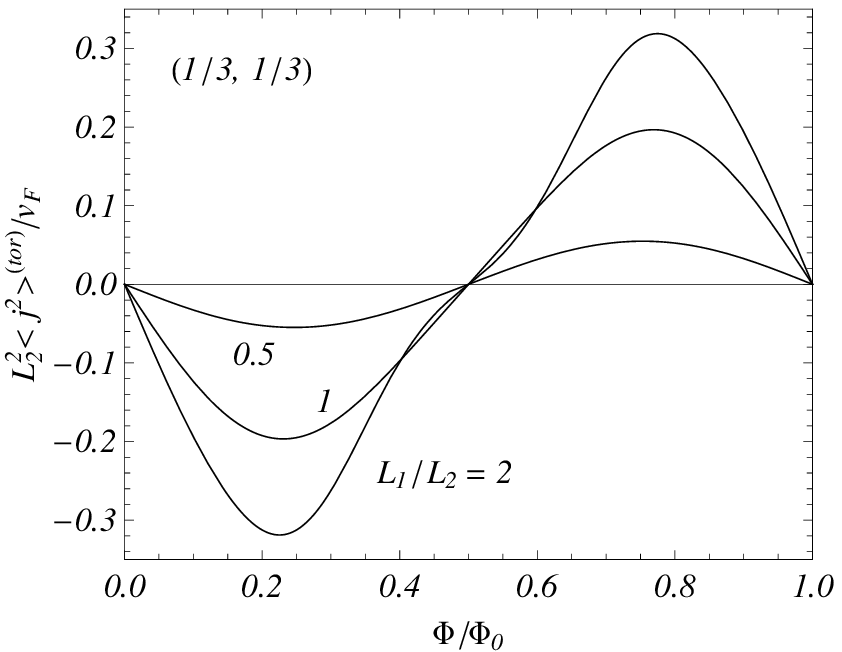,width=7.cm,height=6.cm}%
\end{tabular}%
\end{center}
\caption{The VEV of fermionic current in the massless case for toroidal
nanotubes with $(\protect\alpha _{1},\protect\alpha _{2})=(1/3,0)$ (left
panel) and $(\protect\alpha _{1},\protect\alpha _{2})=(1/3,1/3)$ (right
panel) as a function of the magnetic flux for various values of the ratio $%
L_{1}/L_{2}$.}
\label{fig5}
\end{figure}

The dependence of the fermionic current on the magnetic flux is presented in
figure \ref{fig5} for different values of the ratio $L_{1}/L_{2}$. The left
and right panels correspond to toroidal nanotubes with phases $(\alpha
_{1},\alpha _{2})=(1/3,0)$ and $(1/3,1/3)$, respectively.

\section{Conclusion}

\label{sec:Conc}

We have investigated the VEV of fermionic current for a massive spinor field
in the background of flat spacetime with spatial topology $R^{p}\times S^{q}$%
. Along the compact dimensions the field obeys generic quasiperiodic
boundary conditions (\ref{BC1}). In addition, we have assumed the presence
of a constant gauge field. For the evaluation of the mode sum of the
fermionic current two different approaches have been used. They give two
alternative representations of the vacuum current. In the first approach, we
apply to the mode sum the Abel-Plana type summation formula (\ref{sumform}).
The renormalized VEV of fermionic current components along compact
dimensions is given by formula (\ref{jr1}). The time component and the
components along the uncompactified dimensions vanish. The fermionic current
depends on the phases in the periodicity conditions and on the gauge
potential in the combination (\ref{alfatilde}). It is a periodic function of
the magnetic flux with the period of the flux quantum. In order to obtain an
alternative representation of the vacuum current, in Section \ref{sec:FCZeta}
we have followed the zeta function approach. An exponentially convergent
expression for the analytic continuation of the corresponding mode-sum is
obtained on the basis of the generalized Chowla-Selberg formula. The
corresponding expression for the components of fermionic current along
compact dimensions is given by Eq. (\ref{jr3}). The equivalence of two
representations for the VEV of the fermionic current is directly seen by
using the relation (\ref{Rel3}). As a numerical example, in figure \ref{fig1}
we have depicted the dependence of the vacuum current in the 5-dimensional
Kaluza-Klein model on the phases in the periodicity conditions and on the
mass of the field. In this type models the fermionic current with the
components along compact dimensions is a source of cosmological magnetic
fields.

In Section \ref{sec:FCNano} we gave an application of the general results to
the electrons of a graphene sheet rolled into cylindrical and toroidal
shapes. For the description of relevant low-energy degrees of freedom we
have followed a route based on the effective field theory treatment of
graphene in terms of a pair of Dirac fermions. For this model we have $D=2$
and the topologies $R^{1}\times S^{1}$ and $(S^{1})^{2}$ for cylindrical and
toroidal nanotubes, respectively. Depending on the manner the cylinder is
obtained from the graphene sheet, the phases in the periodicity conditions
for the fields are equal to $0$ for metallic nanotubes and to $\pm 1/3$ for
semiconductor ones. These phases have opposite signs for the two sublattices
of the hexagonal lattice of graphene. In cylindrical nanotubes the total
fermionic current is given by formula (\ref{jrs}). In the absence of
magnetic flux, the total fermionic current vanishes due to the cancellation
of contributions from two sublattices. For toroidal nanotubes the two
equivalent representations for the VEV of fermionic current are given by
Eqs. (\ref{jrtor1n}) and (\ref{jrtor2n}). As in the case of cylindrical
nanotubes, due to the cancellation of contributions coming from separate
sublattices, the fermionic current vanishes in the absence of magnetic flux.
However, the mass terms for two sublattices can be different and in this
case an effective fermionic current appears in the absence of the magnetic
flux.

In a way similar to that used in this paper, we can investigate
the effects of non-trivial topology on the VEV of axial current in
even dimensional spacetimes. It is well-known that in external
electromagnetic and gravitational fields the chiral anomaly
appears in the divergence of the axial current (for a review see
\cite{Trei72}). However, in the problem under consideration these
anomalies are absent as the spacetime is flat and the
electromagnetic field tensor vanishes.

\section*{Acknowledgments}

This work has been supported in part by the EU under the 7th
Framework Program ICT-2007.8.1 FET Proactive 1: Nano-scale ICT
devices and systems Carbon nanotube technology for high-speed
next-generation nano-Interconnects (CATHERINE) project, Grant
Agreement n. 216215. A.A.S. was supported by the Armenian Ministry
of Education and Science Grant No.~119.


\begin{thebibliography}{99}
\bibitem{Lind04} A. Linde, JCAP \textbf{10}, 004 (2004).

\bibitem{Sait98} R. Saito, G. Dresselhaus, and M. S. Dresselhaus, \textit{%
Physical Properties of Carbon Nanotubes} (Imperial College Press, London,
1998); C. Dupas, P. Houdy, and M. Lahmani (Editors), \textit{Nanoscience:
Nanotechnologies and Nanophysics} (Springer, Berlin, 2007).

\bibitem{Seme84} G.W. Semenoff, Phys. Rev. Lett. \textbf{53}, 2449 (1984).

\bibitem{Vinc84} D.P. Di Vincenzo and E.J. Mele, Phys. Rev. B \textbf{29},
1685 (1984); J. Gonz\`{a}lez, F. Guinea, and M.A.H. Vozmediano, Nucl. Phys.
B \textbf{406}, 771 (1993); Phys. Rev. B \textbf{63}, 134421 (2001); H.-W.
Lee and D.S. Novikov, Phys. Rev. B \textbf{68}, 155402 (2003); S.G.
Sharapov, V.P. Gusynin, and H. Beck, Phys. Rev. B \textbf{69}, 075104
(2004); K.S. Novoselov et al, Nature \textbf{438}, 197 (2005); D. S. Novikov
and L. S. Levitov, Phys. Rev. Lett. \textbf{96}, 036402 (2006); E. Perfetto,
J. Gonz\'{a}lez, F. Guinea, S. Bellucci, and P. Onorato, Phys. Rev. B
\textbf{76}, 125430 (2007); A.H. Castro Neto, F. Guinea, N.M.R. Peres, K.S.
Novoselov, and A.K. Geim, Rev. Mod. Phys. \textbf{81}, 109 (2009).

\bibitem{Liu97} H. Liu, H. Dai, J.H. Hafner, D.T. Colbert, R.E. Smalley,
S.J. Tans, and C. Dekker, Nature \textbf{385}, 780 (1997); R. Martel, H.R.
Shea, and P. Avouris, Nature \textbf{398}, 299 (1999).

\bibitem{Most97} V.M. Mostepanenko and N.N. Trunov, \textit{The Casimir
Effect and Its Applications} (Clarendon, Oxford, 1997).

\bibitem{Eliz94} E. Elizalde, S.D. Odintsov, A. Romeo, A.A. Bytsenko and S.
Zerbini, \textit{Zeta regularization techniques with applications} (World
Scientific, Singapore, 1994).

\bibitem{Milt02} K.A. Milton, \textit{The Casimir Effect: Physical
Manifestation of Zero-Point Energy} (World Scientific, Singapore, 2002).

\bibitem{Bord09} M. Bordag, G.L. Klimchitskaya, U. Mohideen, and V.M.
Mostepanenko, \textit{Advances in the Casimir Effect} (Oxford University
Press, Oxford, 2009).

\bibitem{Duff86} M.J. Duff, B.E.W. Nilsson, and C.N. Pope, Phys. Rep.
\textbf{130}, 1 (1986); A.A. Bytsenko, G. Cognola, L. Vanzo, and S. Zerbini,
Phys. Rep. \textbf{266}, 1 (1996).

\bibitem{Klim09} G.L. Klimchitskaya, U. Mohidden, and V.M. Mostepanenko,
Rev. Mod. Phys. \textbf{81}, 1827 (2009).

\bibitem{Milt03} K.A. Milton, Grav. Cosmol. \textbf{9}, 66 (2003); E.
Elizalde, S. Nojiri, and S.D. Odintsov, Phys. Rev D \textbf{70}, 043539
(2004); E. Elizalde, J. Phys. A \textbf{39}, 6299 (2006); B. Greene and J.
Levin, J. High Energy Phys. \textbf{0711}, 096 (2007); P. Burikham, A.
Chatrabhuti, P. Patcharamaneepakorn, and K. Pimsamarn, J. High Energy Phys.
\textbf{0807}, 013 (2008).

\bibitem{CasTor} J.S. Dowker and R. Critchley, J. Phys. A: Math. Gen.
\textbf{9}, 535 (1976); R. Banach and J.S. Dowker, J. Phys. A: Math. Gen.
\textbf{12}, 2545 (1979); B.S. DeWitt, C.F. Hart, and C.J. Isham, Physica A
\textbf{96}, 197 (1979); S.G. Mamayev and N.N. Trunov, Russian Phys. J.
\textbf{22}, 766 (1979); \textbf{23}, 551 (1980); L.H. Ford, Phys. Rev. D
\textbf{21}, 933 (1980); J. Ambj\o rn and S. Wolfram, Ann. Phys. \textbf{147}%
, 1 (1983); S.G. Mamayev and V.M. Mostepanenko, In \textit{Proceedings of
the Third Seminar on Quantum Gravity} (World Scientific, Singapore, 1985);
Yu.P. Goncharov and A.A. Bytsenko, Phys. Lett. B \textbf{160}, 385 (1985);
Yu.P. Goncharov and A.A. Bytsenko, Nucl. Phys. B \textbf{271}, 726 (1986);
Yu.P. Goncharov and A.A. Bytsenko, Class. Quant. Grav. \textbf{4}, 555
(1987); E. Elizalde, Z. Phys. C \textbf{44}, 471 (1989); E. Ponton and E.
Poppitz, JHEP \textbf{0106}, 019 (2001); H. Queiroz, J.C. da Silva, F.C.
Khanna, J.M.C. Malbouisson, M. Revzen, and A.E. Santana, Ann. Phys. \textbf{%
317}, 220 (2005); A.A. Saharian and M.L. Mkhitaryan, Eur. Phys. J. C, in
press, arXiv:0911.1260.

\bibitem{Saha08} A.A. Saharian and M. R. Setare, Phys. Lett. B \textbf{659},
367 (2008); S. Bellucci and A. A. Saharian, Phys. Rev. D \textbf{77}, 124010
(2008).

\bibitem{Saha08f} A.A. Saharian, Class. Quantum Grav. \textbf{25}, 165012
(2008); E.R. Bezerra de Mello and A.A. Saharian, JHEP \textbf{0812}, 081
(2008).

\bibitem{Bell09a} S. Bellucci and A.A. Saharian, Phys. Rev. D \textbf{79},
085019 (2009).

\bibitem{Bell09b} S. Bellucci and A.A. Saharian, Phys. Rev. D \textbf{80},
105003 (2009).

\bibitem{Beze94CosStr} V.B. Bezerra and E.R. Bezerra de Mello, Class.
Quantum Grav. \textbf{11}, 457 (1994); E.R. Bezerra de Mello, Class. Quantum
Grav. \textbf{11}, 1415 (1994); L. Sriramkumar, Class. Quantum Grav. \textbf{%
18}, 1015 (2001); E.R. Bezerra de Mello, arXiv:0907.4139; Yu.A. Sitenko and
N.D. Vlasii, Class. Quantum Grav. \textbf{26}, 195009 (2009).

\bibitem{Beze08} E. R. Bezerra de Mello and A. A. Saharian, Phys. Rev. D
\textbf{78}, 045021 (2008).

\bibitem{Mama76} S.G. Mamayev, V.M. Mostepanenko, and A.A. Starobinsky, Sov.
Phys. JETP \textbf{43}, 823 (1976) [Zh. Eksp. Teor. Fiz. \textbf{70}, 1577
(1976)].

\bibitem{Saha08b} A. A. Saharian, \textit{The Generalized Abel-Plana Formula
with Applications to Bessel Functions and Casimir Effect} (Yerevan State
University Publishing House, Yerevan, 2008); Preprint ICTP/2007/082;
arXiv:0708.1187.

\bibitem{Prud86} A.P. Prudnikov, Yu.A. Brychkov, and O.I. Marichev,
Integrals and Series (Gordon and Breach, New York, 1986), Vol. 2.

\bibitem{Eliz95} E. Elizalde,\textit{\ Ten Physical Applications of Spectral
Zeta Functions}, Lecture Notes in Physics (Springer-Verlag, Berlin, 1995);

\bibitem{Kirs01} K. Kirsten, \textit{Spectral Functions in Mathematics and
Physics} (CRC Press, Boca Raton, FL, 2001).

\bibitem{Eliz98} E. Elizalde, Commun. Math. Phys. \textbf{198}, 83 (1998);
E. Elizalde, J. Phys. A: Math. Gen. \textbf{34}, 3025 (2001).

\bibitem{Cham00} C. Chamon, Phys. Rev. B \textbf{62}, 2806 (2000); C.-Y.
Hou, C. Chamon, and C. Mudry, Phys. Rev. Lett. \textbf{98}, 186809 (2007).

\bibitem{Giov07} G. Giovannetti, P.A. Khomyakov, G. Brocks, P.J. Kelly, and
J. van den Brink, Phys. Rev. B \textbf{76}, 073103 (2007); S.Y. Zhou et al.,
Nature Mater. \textbf{6}, 770 (2007).

\bibitem{Seme08} G.W. Semenoff, V. Semenoff, and F. Zhou, Phys. Rev. Lett.
\textbf{101}, 087204 (2008).

\bibitem{AhBohm} H. Ajiki and T. Ando, J. Phys. Soc. Jpn. \textbf{62}, 1255
(1993); Physica B \textbf{201}, 349 (1994); A. Bachtold et al., Nature
\textbf{397}, 673 (1999); S. Zaric et al., Science \textbf{304}, 1129
(2004); U.S. Coskun et al., Science \textbf{304}, 1132 (2004); J. Cao, Q.
Wang, M. Rolandi, and H. Dai, Phys. Rev. Lett. \textbf{93}, 216803 (2004);
B. Lassagne et al., Phys. Rev. Lett. \textbf{98}, 176802 (2007); M.-G. Kang
et al., Phys. Rev. B \textbf{77}, 113408 (2008).

\bibitem{Sasa02} K. Sasaki, Phys. Lett. A \textbf{296}, 237 (2002); K.
Sasaki, Phys. Rev. B \textbf{65}, 155429 (2002).

\bibitem{Bluh09} H. Bluhm et al., Phys. Rev. Lett. \textbf{102}, 136802
(2009); A.C. Bleszynski-Jayich et al., Science \textbf{326}, 272 (2009).

\bibitem{Lin98} M.F. Lin and D.S. Chuu, Phys. Rev. B \textbf{57}, 6731
(1998); M. Marganska and M. Szopa, Acta Phys. Pol. B \textbf{32}, 427
(2001); S. Latil, S. Roche, and A. Rubio, Phys. Rev. B \textbf{67}, 165420
(2003); R. B. Chen et al., Carbon \textbf{42}, 2837 (2004); K. Sasaki and Y.
Kawazoe, Prog. Theor. Phys. \textbf{112}, 369 (2004); Z. Zhang, J. Yuan, M.
Qiu, J. Peng, and F. Xiao, J. Appl. Phys. \textbf{99}, 104311 (2006); N. Xu,
J.W. Ding, H.B. Chen, and M.M. Ma, Eur. Phys. J. B \textbf{67}, 71 (2009).

\bibitem{Trei72} S.B. Treiman, R. Jackiw, and D.J. Gross, \textit{Lectures
on Current Algebra and Its Applications} (Princeton University Press,
Princeton, 1972); F. Bastianelli and P. Van Nieuwenhuizen, \textit{Path
Integrals and Anomalies in Curved Space} (Cambridge University Press,
Cambridge, 2006).
\end{thebibliography}
\end{document}